\documentclass{article}

\usepackage[in]{fullpage}  
\usepackage{natbib} 
\usepackage{mathpazo} 

\usepackage[utf8]{inputenc} %
\usepackage[T1]{fontenc}    %
\usepackage[hypertexnames=false]{hyperref}  %
\usepackage{url}            %
\usepackage{booktabs}       %
\usepackage{amsfonts}       %
\usepackage{nicefrac}       %
\usepackage{microtype}      %
\usepackage{lipsum}         %

\usepackage{algorithmicx}
\usepackage{algpseudocode}

\usepackage{wrapfig}
\usepackage[table]{xcolor}
\usepackage{makecell}

\usepackage{url}            %
\usepackage{booktabs}       %
\usepackage{multirow}    
\usepackage{amsfonts}       %
\usepackage{nicefrac}       %
\usepackage{microtype}      %
\usepackage{natbib}
\usepackage{enumerate}
\usepackage{hhline}
\usepackage{makecell}
\usepackage{pifont}

\usepackage{graphicx} %
\usepackage{subfigure}
\usepackage{caption}
\usepackage{subcaption}
\usepackage{amsmath}
\usepackage{amsthm}
\usepackage{amssymb}
\usepackage{tikz}
\usepackage{xcolor}
\usetikzlibrary{arrows}

\allowdisplaybreaks

\usepackage{mathrsfs}

\usepackage{algorithm}
\usepackage{hyperref}
\usepackage{bm}

\allowdisplaybreaks

\usepackage[capitalize,noabbrev]{cleveref}
\crefname{thm}{Theorem}{Theorems}
\crefname{lem}{Lemma}{Lemmas}
\crefname{cor}{Corollary}{Corollaries}
\crefname{prop}{Proposition}{Propositions}
\crefname{asmp}{Assumption}{Assumptions}
\crefname{defn}{Definition}{Definitions}
\crefname{oracle}{Oracle}{Oracles}
\crefname{fact}{Fact}{Facts}
\crefname{conj}{Conjecture}{Conjectures}
\crefname{rem}{Remark}{Remarks}
\crefname{example}{Example}{Examples}
\crefname{condition}{Condition}{Conditions}
\crefname{exercise}{Exercise}{Exercises}
\crefname{algorithm}{Algorithm}{Algorithms}
\crefname{table}{Table}{Tables}
\crefname{figure}{Figure}{Figures}
\crefname{section}{Section}{Sections}
\crefname{subsection}{Section}{Sections}
\crefname{appendix}{Appendix}{Appendices}
\crefname{message}{Message}{Messages}

\definecolor{red}{rgb}{1, 0, 0}

\definecolor{green}{rgb}{0, 1, 0}

\definecolor{blue}{rgb}{0, 0, 1}

\definecolor{orange}{rgb}{1, 0.4, 0.0}

\input{math_commands}

\usepackage{tikz}                    % to draw the brace
\usetikzlibrary{decorations.pathreplacing,calc} % brace + coordinate math

\theoremstyle{definition}

\newtheorem{theorem}{Theorem}

\usepackage{soul}

\usepackage{framed} % this is for snugshade
\colorlet{shadecolor}{orange!15}

\AtBeginDocument{\setlength\abovedisplayskip{5pt}}
\AtBeginDocument{\setlength\belowdisplayskip{5pt}}

\hypersetup{
    colorlinks=true,
    citecolor=blue,
    linkcolor=blue,
}

\usepackage{listings} %

\definecolor{codegreen}{rgb}{0,0.6,0}
\definecolor{codegray}{rgb}{0.5,0.5,0.5}
\definecolor{codepurple}{rgb}{0.58,0,0.82}
\definecolor{codeblue}{rgb}{0,0,1}
\definecolor{backcolour}{rgb}{0.95,0.95,0.92}
\definecolor{key-color}{rgb}{0.8, 0.47, 0.196}

\lstdefinestyle{mystyle}{
    backgroundcolor=\color{backcolour},   
    commentstyle=\color{codegreen},
    numberstyle=\tiny\color{codegray},
    stringstyle=\color{codepurple},
    basicstyle=\ttfamily\footnotesize,
    breakatwhitespace=false,         
    breaklines=true,                 
    captionpos=b,                    
    keepspaces=true,                 
    numbers=left,                    
    numbersep=5pt,                  
    showspaces=false,                
    showstringspaces=false,
    showtabs=false,                  
    tabsize=2,
    language=Python,
    emph={lm},
    emphstyle={\color{blue}},
    classoffset=1, %
    otherkeywords={sum},
    morekeywords={rm, mean},
    keywordstyle=\color{codegreen},
    classoffset=0,
}
\newcommand{\yushunrevise}[1]{\textcolor{red}{ #1}}

\title{$XX^{t}$ Can Be Faster}

\makeatletter
\def\@fnsymbol#1{\ensuremath{\ifcase#1\or *\or \dagger\or \ddagger\or
   \mathsection\or \sharp\or \Diamond\or \mathparagraph\or \|\or
   \or \ddagger\ddagger \else\@ctrerr\fi}}
\makeatother

\usepackage{authblk}
\author[1,2]{Dmitry Rybin \thanks{Correspondence author. Email: \texttt{dmitryrybin@link.cuhk.edu.cn}.}}
\author[1,2]{Yushun Zhang \thanks{Email: \texttt{yushunzhang@link.cuhk.edu.cn}.}}

\author[1,2]{Zhi-Quan Luo\thanks{Email: \texttt{luozq@cuhk.edu.cn}}}
\affil[1]{The Chinese University of Hong Kong, Shenzhen, China}
\affil[2]{Shenzhen Research Institute of Big Data}

\begin{document}

\maketitle
\vspace{-0.5cm}
\begin{abstract}
We present RXTX, a new algorithm for computing the product of matrix by its transpose $XX^{t}$ for $X\in \mathbb{R}^{n\times m}$.  RXTX uses 5\% fewer multiplications and 5\% fewer operations (additions and multiplications) than State-of-the-Art algorithms. Note that the accelerations not only holds asymptotically for large matrices with $n \rightarrow \infty$, but also for small matrices including $n = 4$. The algorithm was discovered by combining Machine Learning-based search methods with Combinatorial Optimization.
\end{abstract}

\vspace{-0.5cm}

\section{Introduction}
\label{introduction}

\setlength{\tabcolsep}{3pt}
\begin{table}[H]
    \centering
    \resizebox{0.98\linewidth}{!}{%
    \begin{tabular}{|c|c|c|}
    \hline 
       Algorithm & Previous State-of-the-Art for $XX^{t}$ & RXTX \\
        \hline
      \makecell{Illustration in \\ matrix form} & \hspace{-2.3ex} 
$\begin{array}{cc}
  & \left(\begin{array}{>{\columncolor{red!15}}c >{\columncolor{yellow!15}}c}  \hspace*{0.67cm} A^{t} \hspace*{0.67cm}  & \hspace*{0.67cm} C^{t} \hspace*{0.67cm} \\ B^{t} & D^{t} \end{array}\right) \\
\\[-2pt] 
\left(\begin{array}{cc}
\rowcolor{red!15} A & B \\ 

\rowcolor{yellow!15} C & D 
\end{array}\right) & \left(\begin{array}{cc} \cellcolor{teal!15} AA^{t} + BB^{t} & \cellcolor{violet!15} AC^{t} + BD^{t} \\ * & \cellcolor{teal!15} CC^{t} + DD^{t} \end{array}\right)
\end{array}$ \hspace{-2.3ex}
 & \hspace{-2.2ex} $ 
\begin{array}{cc}
  & \left(\begin{array}{>{\columncolor{red!15}}c >{\columncolor{yellow!15}}c >{\columncolor{green!15}}c >{\columncolor{blue!15}}c} 
   \hspace*{0.14cm} & \hspace*{0.14cm} & \hspace*{0.14cm} & \hspace*{0.14cm} \\ 
   & & & \\ 
   & & & \\ 
   & & &  \end{array}\right) \\
\\
\left(\begin{array}{cccc}
\rowcolor{red!15}  \hspace*{0.14cm} & \hspace*{0.14cm} & \hspace*{0.14cm} & \hspace*{0.14cm} \\ 
\rowcolor{yellow!15}  & & & \\
\rowcolor{green!15}  & & & \\
\rowcolor{blue!15}  & & & \\ 
\end{array}\right) & \left(\begin{array}{cccc} \cellcolor{teal!15} & \cellcolor{violet!15} & \cellcolor{violet!15} & \cellcolor{violet!15} \\
*  & \cellcolor{violet!15} & \cellcolor{violet!15} & \cellcolor{violet!15}\\
* & * & \cellcolor{violet!15} & \cellcolor{violet!15} \\
* & * & * & \hspace*{0.08cm} \cellcolor{teal!15} \end{array}\right)
\end{array} 
$ \hspace{-2.2ex} \\

         \hline
       Recursive expression & $S(n) = $\colorbox{teal!15}{$4S(n/2)$} + \colorbox{violet!15}{$2M(n/2)$} & $R(n) = $\colorbox{teal!15}{$8R(n/4)$} $+$ \colorbox{violet!15}{$26M(n/4)$} \\
         \hline
      \makecell{Asymptotic speedup  \\ (\# multiplications for $n\rightarrow \infty$)} & $S(n)\sim \frac{2}{3}M(n)$ & $R(n)\sim \frac{26}{41}M(n)$ \yushunrevise{$({\bf 5\% }\downarrow)$} \\
        \hline
      \makecell{Non-asymptotic speedup  \\ (\# multiplications for $n=4$)}& $38$ & $34$ \yushunrevise{$({\bf 10\% }\downarrow)$}  \\
        \hline
    \end{tabular}
    }
    \caption{Comparison between the proposed algorithm RXTX and previous State-of-the-Art (SotA) algorithm for computing $XX^{t}$ for $X\in \mathbb{R}^{n\times m}$, $n,m \geq 4$. RXTX is based on recursive $4\times 4$ block matrix multiplication. It uses $8$ recursive calls and $26$ general products. The previous SotA uses $16$ recursive calls and $24$ general products. $R(n), S(n), M(n)$ - are the number of multiplications performed by RXTX, previous SotA, and Strassen algorithm respectively for $X\in \mathbb{R}^{n\times m}$. The asymptotic constant of RXTX, $\frac{26}{41} \approx 0.6341$, is approximately 5\% smaller than that of the previous state-of-the-art, $\frac{2}{3} \approx 0.6666$.}
\end{table}
\setlength{\tabcolsep}{6pt}

% RXTX asymptotic constant $26/41\approx 0.6341$ is $5\%$ smaller than $2/3 \approx 0.6666$, which is asymptotic constant of previous SotA.

\begin{algorithm}
\caption{RXTX - new algorithm for $XX^{t}$}\label{alg:rxtx}
\begin{algorithmic}[1]
\State \textbf{Input:} $4 \times 4$ block-matrix $X$
\State \textbf{Output:} $C = XX^{t}$ using $8$ recursive calls and $26$ general products. 
\begingroup
    \setlength{\fboxsep}{0.8pt} 
\State $m_1 = \colorbox{red!15}{$(-X_2 + X_3 - X_4 + X_8)$} \cdot \colorbox{green!15}{$(X_{8} + X_{11})^{t}$}$ \tikz[remember picture] \coordinate (brace2-start);
\State $m_2 = \colorbox{red!15}{$(X_{1} - X_{5} - X_{6} + X_{7})$} \cdot \colorbox{green!15}{$(X_{15} + X_{5})^{t}$}$
\State $m_3 = \colorbox{red!15}{$(-X_{2} + X_{12})$} \cdot \colorbox{green!15}{$(-X_{10} + X_{16} + X_{12})^{t}$}$
\State $m_4 = \colorbox{red!15}{$(X_{9} - X_{6})$} \cdot \colorbox{green!15}{$(X_{13} + X_{9} - X_{14})^{t}$}$
\State $m_5 = \colorbox{red!15}{$(X_{2} + X_{11})$} \cdot \colorbox{green!15}{$(-X_{6} + X_{15} - X_{7})^{t}$}$
\State $m_6 = \colorbox{red!15}{$(X_{6} + X_{11})$} \cdot \colorbox{green!15}{$(X_{6} + X_{7} - X_{11})^{t}$}$
\State $m_7 =  \colorbox{red!15}{$X_{11}$} \cdot \colorbox{green!15}{$(X_{6} + X_{7})^{t}$}$
\State $m_8 = \colorbox{red!15}{$X_{2}$} \cdot \colorbox{green!15}{$(-X_{14} - X_{10} + X_{6} - X_{15} + X_{7} + X_{16} + X_{12})^{t}$}$
\State $m_9 = \colorbox{red!15}{$X_{6}$} \cdot \colorbox{green!15}{$(X_{13} + X_{9} - X_{14} - X_{10} + X_{6} + X_{7} - X_{11})^{t}$}$
\State $m_{10} = \colorbox{red!15}{$(X_{2} - X_{3} + X_{7} + X_{11} + X_{4} - X_{8})$} \cdot \colorbox{green!15}{$X_{11}^{t}$}$
\State $m_{11} = \colorbox{red!15}{$(X_{5} + X_{6} - X_{7})$} \cdot \colorbox{green!15}{$X_{5}^{t}$}$
\State $m_{12} = \colorbox{red!15}{$(X_{2} - X_{3} + X_{4})$} \cdot \colorbox{green!15}{$X_{8}^{t}$}$
\State $m_{13} = \colorbox{red!15}{$(-X_{1} +X_{5} + X_{6} + X_{3} - X_{7} + X_{11})$} \cdot \colorbox{green!15}{$X_{15}^{t}$}$
\State $m_{14} = \colorbox{red!15}{$(-X_{1} + X_{5} + X_{6})$} \cdot \colorbox{green!15}{$(X_{13} + X_{9} + X_{15})^{t}$}$
\State $m_{15} = \colorbox{red!15}{$(X_{2} + X_{4} - X_{8})$} \cdot \colorbox{green!15}{$(X_{11} + X_{16} + X_{12})^{t}$}$
\State $m_{16} = \colorbox{red!15}{$(X_{1} - X_{8})$} \cdot \colorbox{green!15}{$(X_{9} - X_{16})^{t}$}$
\State $m_{17} = \colorbox{red!15}{$X_{12}$} \cdot \colorbox{green!15}{$(X_{10} - X_{12})^{t}$}$
\State $m_{18} = \colorbox{red!15}{$X_{9}$} \cdot \colorbox{green!15}{$(X_{13} - X_{14})^{t}$}$
\State $m_{19} = \colorbox{red!15}{$(-X_{2} + X_{3})$} \cdot \colorbox{green!15}{$(-X_{15} + X_{7} + X_{8})^{t}$}$
\State $m_{20} = \colorbox{red!15}{$(X_{5} + X_{9} - X_{8})$} \cdot \colorbox{green!15}{$X_{9}^{t}$}$
\State $m_{21} = \colorbox{red!15}{$X_{8}$} \cdot \colorbox{green!15}{$(X_{9} - X_{8} + X_{12})^{t}$}$
\State $m_{22} = \colorbox{red!15}{$(-X_{6} + X_{7})$} \cdot \colorbox{green!15}{$(X_{5} + X_{7} - X_{11})^{t}$}$
\State $m_{23} = \colorbox{red!15}{$X_{1}$} \cdot \colorbox{green!15}{$(X_{13} - X_{5} + X_{16})^{t}$}$
\State $m_{24} = \colorbox{red!15}{$(-X_{1} + X_{4} + X_{12})$} \cdot \colorbox{green!15}{$X_{16}^{t}$}$
\State $m_{25} = \colorbox{red!15}{$(X_{9} + X_{2} + X_{10})$} \cdot \colorbox{green!15}{$X_{14}^{t}$}$
\State $m_{26} = \colorbox{red!15}{$(X_{6} + X_{10} + X_{12})$} \cdot \colorbox{green!15}{$X_{10}^{t}$}$ \tikz[remember picture] \coordinate (brace2-end);

\begin{tikzpicture}[overlay, remember picture]
  \draw[decorate, decoration={brace,amplitude=6pt}, % ← curly brace style
    xshift=0.9em                       % ← push it a little to the right
  ] 
    ($(brace2-start.east)+(+12.2ex,+1.4ex)$) --   % and ends a hair below top line
    ($(brace2-end.east)+(+23.9ex,-0.4ex)$)   % path starts a hair above bottom line
    
    node[midway,xshift=5.5em]{{\bf 26 multiplications}}; % label
\end{tikzpicture}\vspace*{\dimexpr -\baselineskip + 0.3ex\relax}

\State $s_1 = \colorbox{red!15}{$X_{1}$}\cdot\colorbox{green!15}{$X_{1}^{t}$}$ \tikz[remember picture] \coordinate (brace-start);
\State $s_2 = \colorbox{red!15}{$X_{2}$}\cdot\colorbox{green!15}{$X_{2}^{t}$}$
\State $s_3 = \colorbox{red!15}{$X_{3}$}\cdot\colorbox{green!15}{$X_{3}^{t}$}$
\State $s_4 = \colorbox{red!15}{$X_{4}$}\cdot\colorbox{green!15}{$X_{4}^{t}$}$
\State $s_5 = \colorbox{red!15}{$X_{13}$}\cdot\colorbox{green!15}{$X_{13}^{t}$}$
\State $s_6 = \colorbox{red!15}{$X_{14}$}\cdot\colorbox{green!15}{$X_{14}^{t}$}$
\State $s_7 = \colorbox{red!15}{$X_{15}$}\cdot\colorbox{green!15}{$X_{15}^{t}$}$
\State $s_8 = \colorbox{red!15}{$X_{16}$}\cdot\colorbox{green!15}{$X_{16}^{t}$}$ \tikz[remember picture] \coordinate (brace-end);

\begin{tikzpicture}[overlay,remember picture]
  \draw[
    decorate,
    decoration={brace,amplitude=6pt}, % ← curly brace style
    xshift=0.9em                       % ← push it a little to the right
  ] 
    ($(brace-start.east)+(+40.2ex,+1.4ex)$) --   % and ends a hair below top line
    ($(brace-end.east)+(+38.6ex,-0.4ex)$)   % path starts a hair above bottom line
    
    node[midway,xshift=5.5em]{{\bf 8 recursive calls}}; % label
\end{tikzpicture}\vspace*{\dimexpr -\baselineskip + 0.3ex\relax}
\State $C_{11} = \colorbox{teal!15}{$s_1 + s_2 + s_3 + s_4$}$
\State $C_{12} = \colorbox{violet!15}{$m_2 - m_5 - m_7 + m_{11} + m_{12} + m_{13} + m_{19}$}$
\State $C_{13} = \colorbox{violet!15}{$m_1 + m_3 + m_{12} + m_{15} + m_{16} + m_{17} + m_{21} - m_{24}$}$
\State $C_{14} = \colorbox{violet!15}{$m_2 - m_3 - m_5 - m_7 - m_8 + m_{11} + m_{13} - m_{17} + m_{23} + m_{24}$}$
\State $C_{22} = \colorbox{violet!15}{$m_1 + m_6 - m_7 + m_{10} + m_{11} + m_{12} + m_{22}$}$
\State $C_{23} = \colorbox{violet!15}{$m_{1} - m_{4} + m_{6} - m_{7} - m_{9} + m_{10} + m_{12} + m_{18} + m_{20} + m_{21}$}$
\State $C_{24} = \colorbox{violet!15}{$m_2 + m_4 + m_{11} + m_{14} + m_{16} - m_{18} - m_{20} + m_{23}$}$
\State $C_{33} = \colorbox{violet!15}{$m_4 - m_6 + m_7 + m_9 - m_{17} - m_{18} + m_{26}$}$
\State $C_{34} = \colorbox{violet!15}{$m_3 + m_5 + m_7 + m_8 + m_{17} + m_{18} + m_{25}$}$
\State $C_{44} = \colorbox{teal!15}{$s_5 + s_6 + s_7 + s_8$}$
\endgroup
\State \Return $C$
\end{algorithmic}
\end{algorithm}

Finding faster matrix multiplication algorithms is a central challenge in computer science and numerical linear algebra. Since the groundbreaking results of \citep{strassen1969} and \citep{winograd1968}, which demonstrated that the number of multiplications required for a general matrix product $AB$ can be significantly reduced, extensive research has emerged exploring this problem. Techniques in the area range from gradient descent approaches \citep{smirnov2013} and heuristics \citep{drevet2011}, to group-theoretic methods \citep{ye2018}, graph-based random walks \citep{kauers2022}, and deep reinforcement learning \citep{fawzi2022}.

Despite this progress, much less attention has been paid to matrix products with \textbf{additional structure}, such as $B = A$ or $B = A^{t}$, or products involving sparsity or symmetry \citep{dumas2020, dumas2023, arrigoni2021}. This is surprising given that expressions like $AA^{t}$ are widely used in fields such as statistics, data analysis, deep learning, finance, and wireless communications. For example, $AA^{t}$ often represents a covariance matrix or Gram matrix, while in linear regression, the solution for the data pair $(X, y)$ involves the data covariance matrix $X^{t}X$:
$$\beta = (X^{t}X)^{-1}X^{t}y.$$
Furthermore, the expression $XX^{t}$ is used in modern algorithms for LLM training. Newton-Schulz step in Muon \citep{jordan2024muon} and update rule in SOAP, Shampoo \citep{soap, shampoo} repeat the $XX^{t}$ computation many times.

In this work, we focus on accelerating matrix product with the special structure $XX^t$, where $X\in \mathbb{R}^{n \times m}$ for general $n,m \geq 4$. From a theoretical standpoint, computing $XX^{t}$ has the same asymptotic complexity as general matrix multiplication, so only constant-factor speedups are possible. We then propose a new algorithm called Recursive $X^tX$ (RXTX), which achieves 5\% constant-factor speedup by exploiting structure specific to $XX^{t}$.  RXTX
was discovered by combining Machine Learning-based search methods with Combinatorial Optimization. The complete form of RXTX is presented in Algorithm \ref{alg:rxtx}.

% From a theoretical standpoint, computing $XX^{t}$ has the same asymptotic complexity as general matrix multiplication. As a result, only constant-factor speedups are possible. The RXTX algorithm, presented in Algorithm \ref{alg:rxtx}, achieves such a speedup by exploiting structure specific to $XX^{t}$.

\subsection{Related Works}
Prior works \citep{ye2016, ye2018} used representation theory and the Cohn–Umans framework to derive new multiplication schemes for structured matrix products. Reinforcement learning methods have also been applied to this domain. For instance, \citet{fawzi2022} used deep RL to compute tensor ranks and discover novel multiplication algorithms. Neural Networks wtih proper training setup can rediscover Strassen and Laderman algorithms for small matrices \citep{elser2016}.

More recently, \cite{dumas2020, dumas2023} proposed optimized schemes for computing $XX^{t}$ over finite fields and complex numbers. To the best of our knowledge, the current state-of-the-art approach for real-valued $XX^{t}$ is due to \citep{arrigoni2021}, which recursively applies Strassen’s algorithm to $2\times 2$ block matrices, reducing the problem to general matrix multiplication. In contrast, our approach uses the structure of $XX^{t}$ in a new way.

\section{Analysis of RXTX}
\label{covariance}

We define
\begin{itemize}
    \item $R(n)$ - number of multiplications performed by RXTX for $n\times n$ matrix
    \item $S(n)$ - number of multiplications performed by recursive Strassen \citep{arrigoni2021} for $n\times n$ matrix
    \item $M(n)$ - number of multiplications performed by Strassen-Winograd algorithm for general product of $n\times n$ matrices
    \item $R_{+}(n)$ - number of additions and multiplications performed by RXTX for $n\times n$ matrix
    \item $S_{+}(n)$ - number of additions and multiplications performed by recursive Strassen \citep{arrigoni2021} for $n\times n$ matrix
    \item $M_{+}(n)$ - number of additions and multiplications performed by Strassen-Winograd algorithm for general product of $n\times n$ matrices
\end{itemize}
The superscript $^{opt}$ indicates an optimal cutoff: for sufficiently small matrices, standard matrix multiplication is used instead of further recursive calls.

\subsection{Number of multiplications}
\begin{theorem} The number of multiplications for RXTX: 
$$R(n) = \frac{26}{41}M(n) + \frac{15}{41}n^{3/2} = \frac{26}{41}n^{\log_{2}7} + \frac{15}{41}n^{3/2}.$$
The number of multiplications for recursive Strassen: 
$$S(n) = \frac{2}{3}M(n) + \frac{1}{3}n^{2} = \frac{2}{3}n^{\log_{2} 7} + \frac{1}{3}n^{2}.$$
\end{theorem}
\begin{proof}
The definition of RXTX involves $8$ recursive calls and $26$ general matrix multiplications. It follows that
$$R(n) = 8R(n/4) + 26 M(n/4).$$
The general solutions to this recursive equation has a form \citep{cormen2009}
$$R(n) = \alpha M(n) + \beta n^{3/2}.$$
Using initial conditions for $n = 1$ and $n = 4$ we find
$$\alpha = \frac{26}{41} \approx 0.6341, \qquad \beta = \frac{15}{41} \approx 0.3658.$$
Similarly, recursive Strassen for $XX^{t}$ uses 4 recursive calls and 2 general matrix multiplications:
$$S(n) = 4S(n/2) + 2M(n/2).$$
General solution form
$$S(n) = \gamma M(n) + \delta n^{2}.$$
Using initial conditions for $n = 1$ and $n = 2$ we find $\gamma = 2/3 \approx 0.6666$ and $\delta = 1/3 \approx 0.3333$.
\end{proof}

\begin{figure}[H]
  \centering
  \begin{minipage}[t]{0.49\textwidth}
    \centering
    \includegraphics[width=\linewidth]{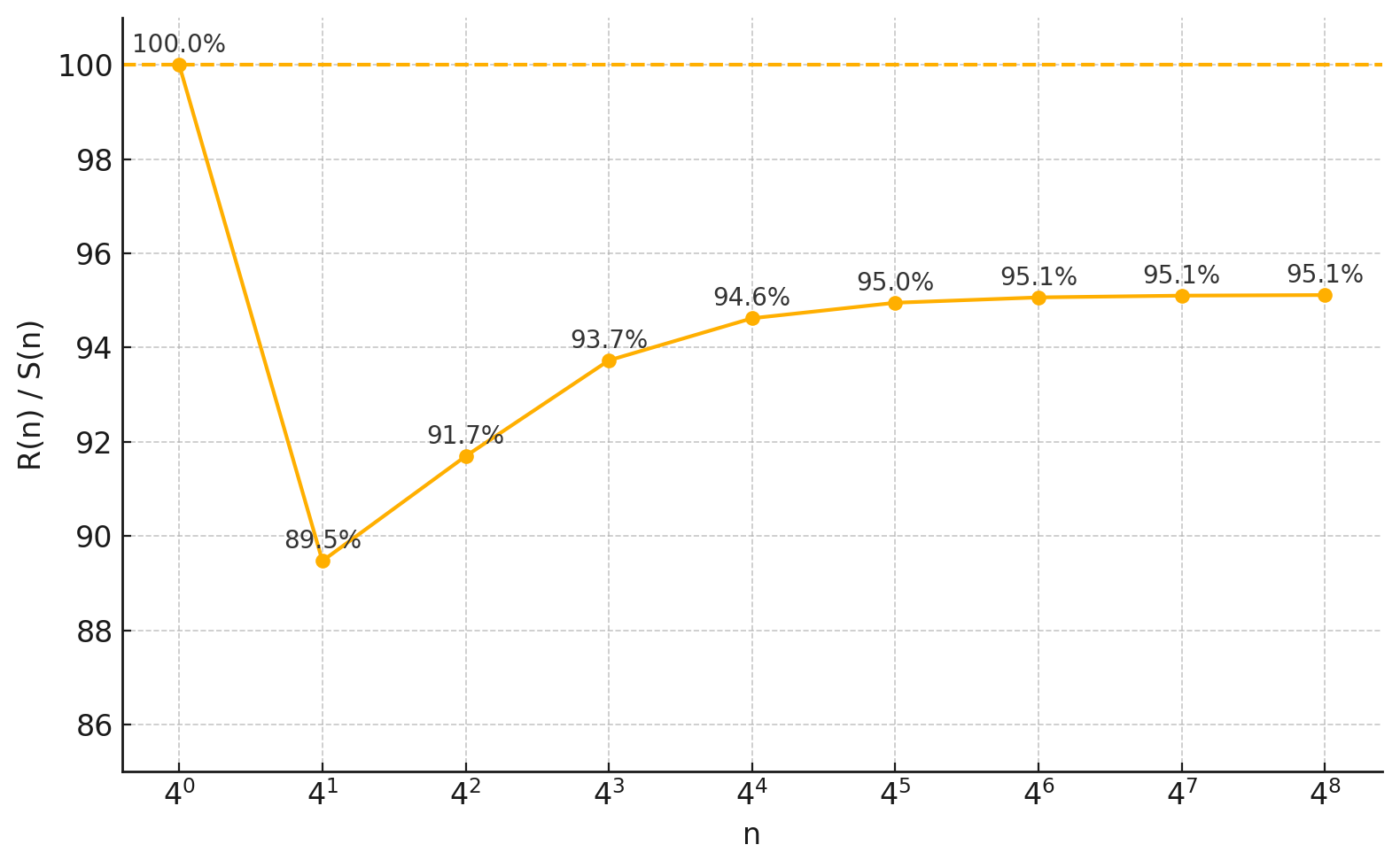}
    Ratio of $R(n)$ to $S(n)$.
  \end{minipage}
  \hfill
  \begin{minipage}[t]{0.49\textwidth}
    \centering
    \includegraphics[width=\linewidth]{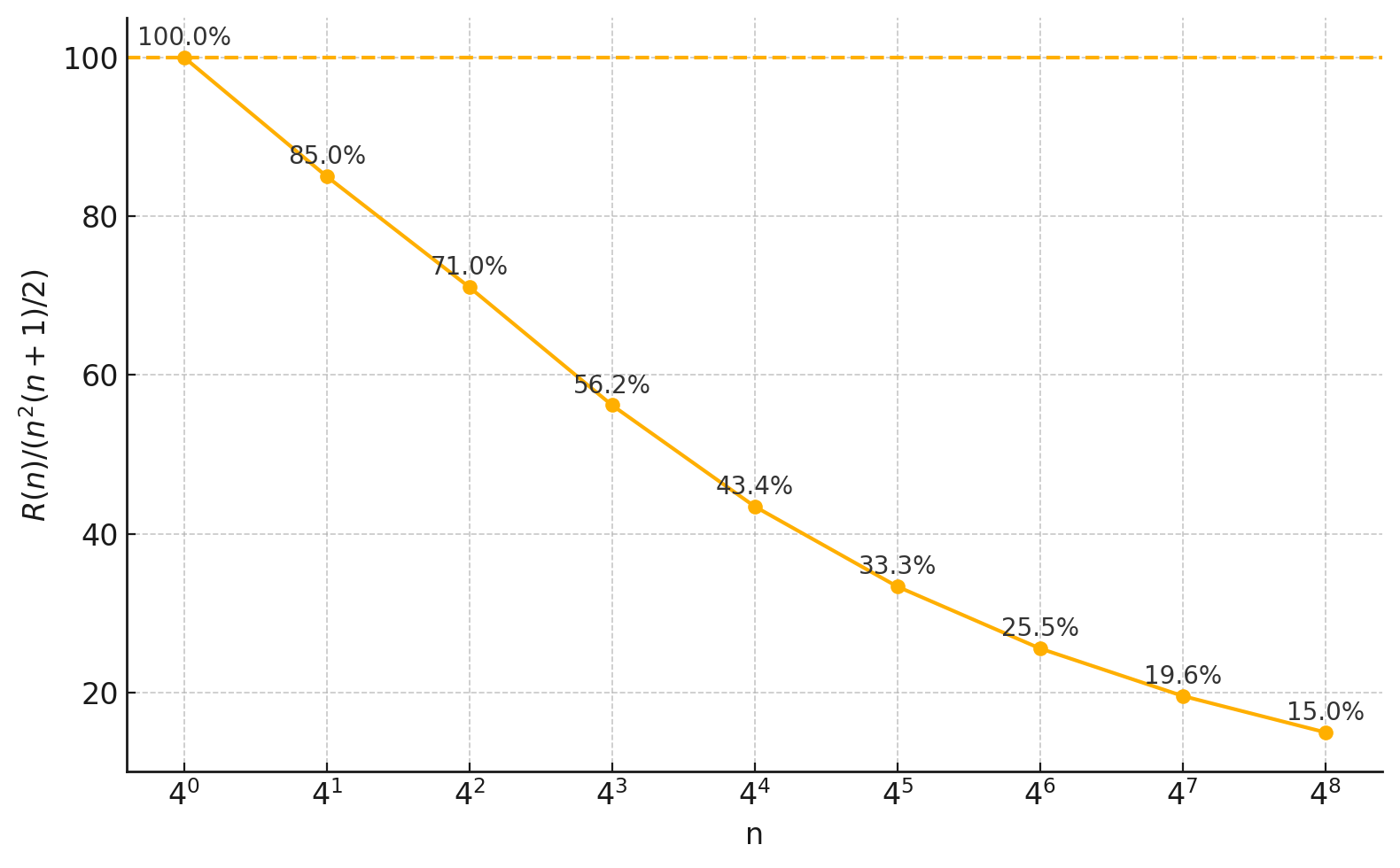}
    Ratio of $R(n)$ to naive algorithm $n^{2}(n+1)/2$.
  \end{minipage}
  \caption{Comparison of number of multiplications of RXTX to previous SotA and naive algorithm.}
  \label{fig:multiplications}
\end{figure}

\begin{figure}[H]
  \centering
  \begin{minipage}[t]{0.49\textwidth}
    \centering
    \includegraphics[width=\linewidth]{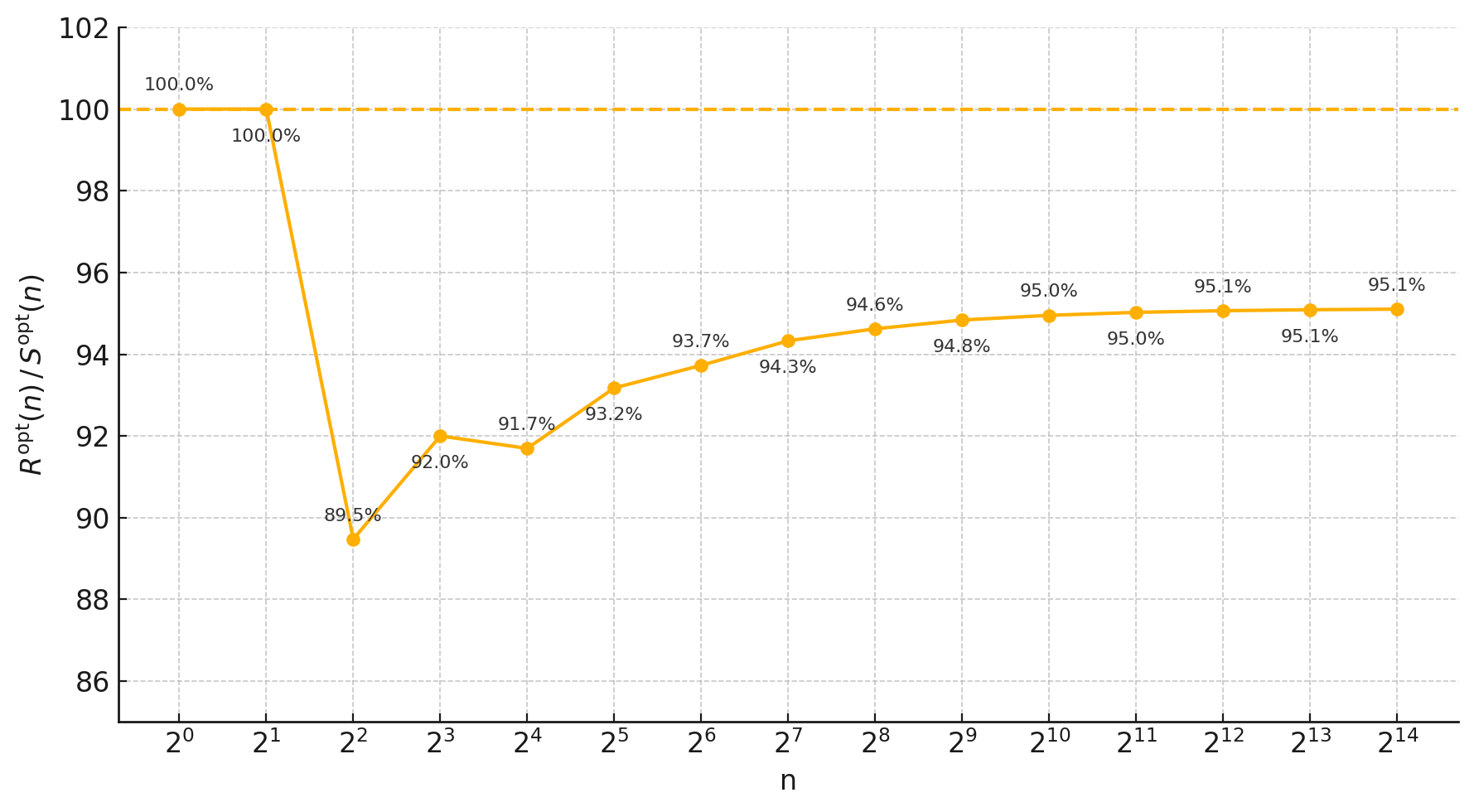}
    Ratio of $R^{opt}(n)$ to $S^{opt}(n)$.
  \end{minipage}
  \hfill
  \begin{minipage}[t]{0.49\textwidth}
    \centering
    \includegraphics[width=\linewidth]{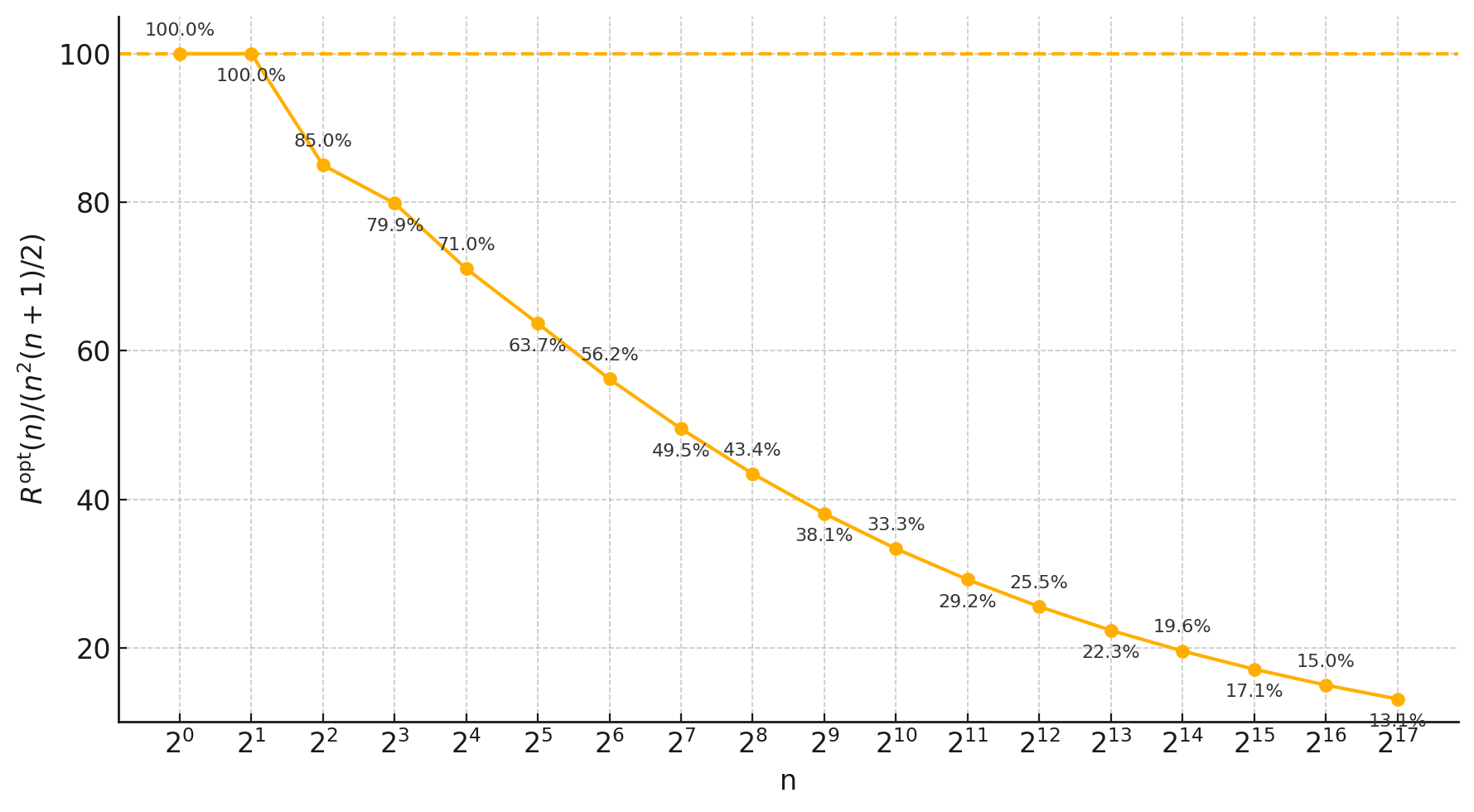}
    Ratio of $R^{opt}(n)$ to $n^{2}(n+1)/2$.
  \end{minipage}
  \caption{Comparison of number of multiplications of RXTX with optimal cutoff to previous SotA and naive algorithm.}
  \label{fig:multiplications_opt}
\end{figure}

In Figure \ref{fig:multiplications} we can see the ratio $R(n)/S(n)$ for $n$ given by powers of $4$. The ratio always stays below $100\%$ and approaches the asymptotic $95\%$, which indicates a $5\%$ reduction in the number of multiplications. Same happens in Figure \ref{fig:multiplications_opt}, where we use optimal cutoff i.e. for small enough matrix sizes we use standard matrix multiplication instead of further recursive calls.

\subsection{Total number of operations}
\begin{theorem}
Total number of additions and multiplications for RXTX:
$$R_{+}(n) = \frac{156}{41}n^{\log_{2}7} - \frac{615}{164}n^{2} + \frac{155}{164}n^{3/2}.$$
Total number of additions and multiplications for recursive Strassen:
$$S_{+}(n) = 4n^{\log_{2}7} -\frac{7}{4}n^{2}\log_{2}n - 3n^{2}.$$
\end{theorem}
\begin{proof}
The definition of RXTX involves $139$ additions of $(n/4) \times (n/4)$ matrices. There exist methods in the literature \citep{martensson2024} to reduce this number by utilizing common sub-expressions that appear in the algorithm \ref{alg:rxtx} e.g. $X_6 + X_7$. For example, while Strassen algorithm uses $18$ additions, its Winograd variant uses only $15$ additions. We designed a custom search that allowed us to reduce the number of additions in RXTX from $139$ to $100$. We provide the resulting addition scheme in Algorithm \ref{alg:stage1} and Algorithm \ref{alg:stage2}. Assuming $100$ additions, we get the recursion
$$R_{+}(n) = 8 R_{+}(n/4) + 26M_{+}(n/4) + 100 (n/4)^2.$$
General solution has a form
$$R_{+}(n) = \frac{26}{41} M_{+}(n) + \alpha n^{2} + \beta n^{3/2}.$$
Plugging the value $n = 1$ and $n = 4$ gives
$$\frac{26}{41} + \alpha + \beta = 1.$$
$$ \frac{26}{41} \cdot 214 + 16\alpha + 8\beta = 134$$
We conclude that 
$$\alpha = -\frac{95}{164} \approx - 0.5793, \qquad \beta = \frac{155}{164} \approx 0.9451.$$
Similarly, definition of recursive Strassen gives
$$S_{+}(n) = 4S_{+}(n/2) + 2M_{+}(n) + 3(n/2)^2.$$
Which has a solution of the form
$$S_{+}(n) = \frac{2}{3}M_{+}(n) + \gamma n^2 \log_{2} n + \delta n^{2}.$$
Plugging values $n=1$ and $n=2$ gives $\gamma = -7/4$ and $\delta = 1/3$. It is known by \cite{cenk2017} that
$$M_{+}(n) = 6n^{\log_{2}7} - 5n^2.$$
It follows that
$$R_{+}(n) = \frac{26}{41}(6n^{\log_{2}7} - 5n^{2}) - \frac{95}{164}n^2 + \frac{155}{164}n^{3/2} = \frac{156}{41} n^{\log_{2}7} - \frac{615}{164}n^{2} + \frac{155}{164}n^{3/2}$$
and
$$S_{+}(n) = \frac{2}{3}(6n^{\log_{2}7} - 5n^{2}) - \frac{7}{4}n^{2}\log_{2} n + \frac{1}{3}n^{2} = 4n^{\log_{2}7} -\frac{7}{4}n^{2}\log_{2}n - 3n^{2}.$$
\end{proof}

\begin{figure}[H]
  \centering
  \begin{minipage}[t]{0.49\textwidth}
    \centering
    \includegraphics[width=\linewidth]{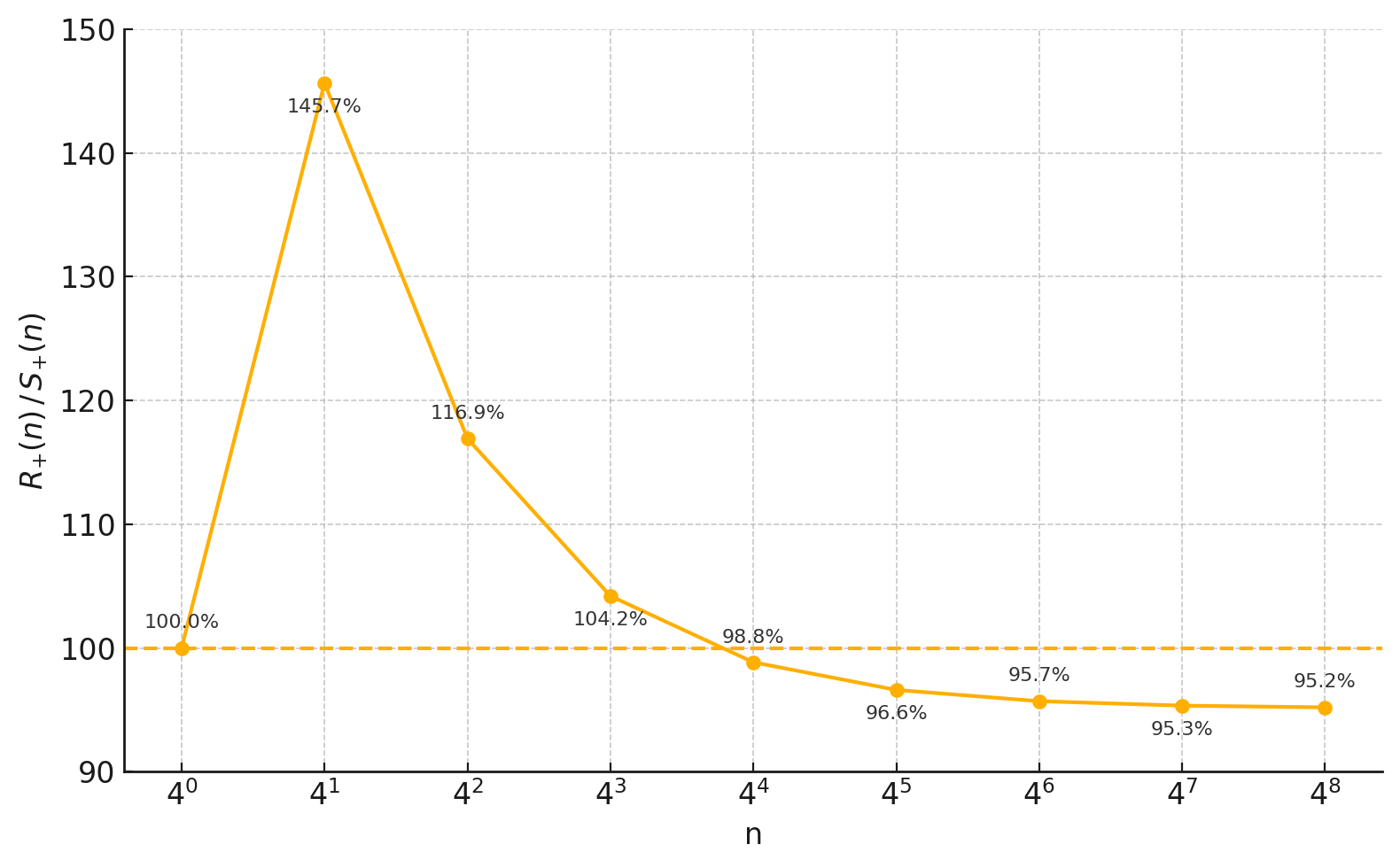}
    Ratio of $R_{+}(n)$ to $S_{+}(n)$.
  \end{minipage}
  \hfill
  \begin{minipage}[t]{0.49\textwidth}
    \centering
    \includegraphics[width=\linewidth]{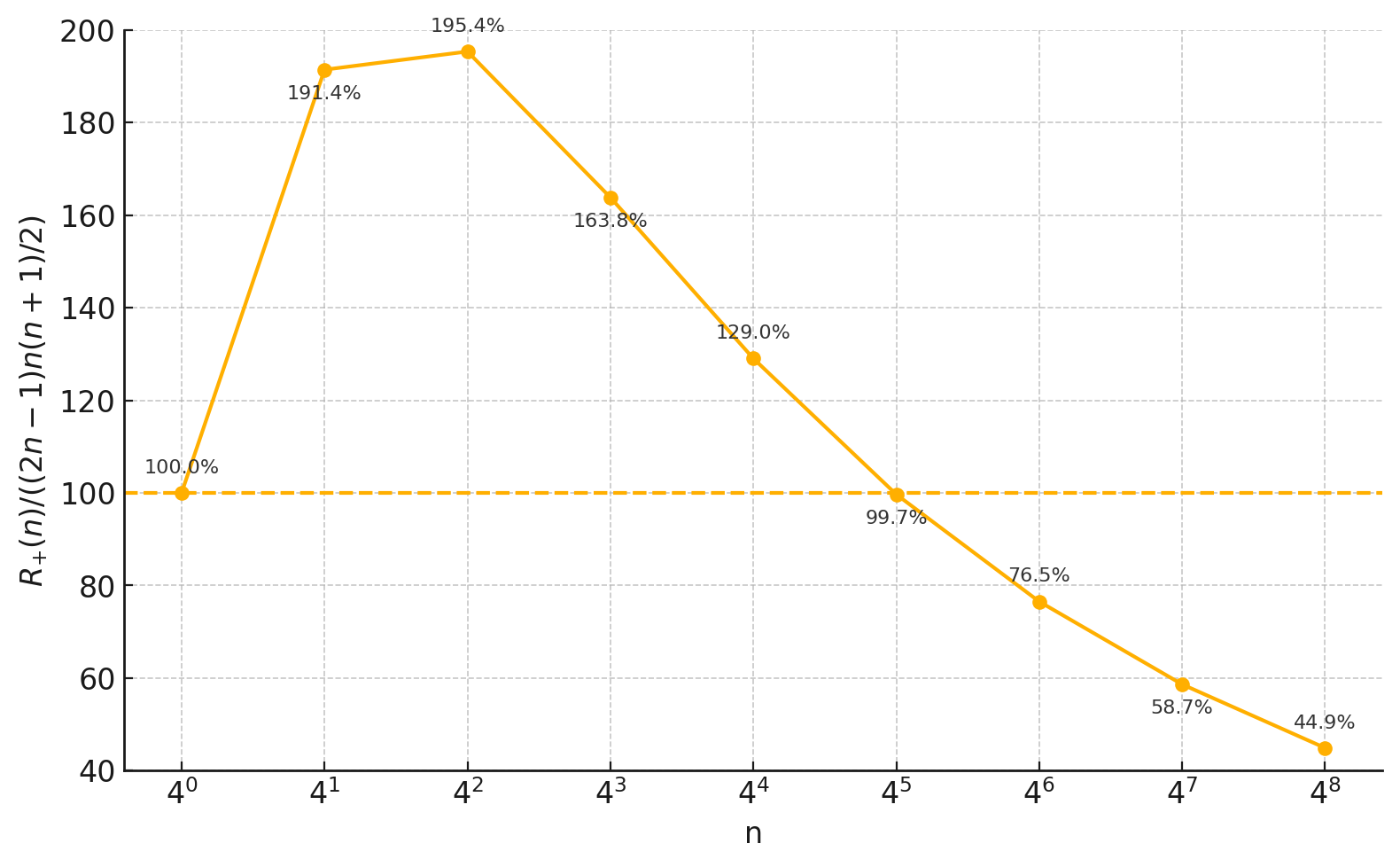}
    Ratio of $R_{+}(n)$ to $(2n-1)n(n+1)/2$.
  \end{minipage}
  \caption{Comparison of number of operations of RXTX to recursive Strassen and naive algorithm. RXTX outperforms recursive Strassen for $n \geq 256$ and naive algorithm for $n \geq 1024$.}
  \label{fig:operations}
\end{figure}

\begin{figure}[H]
  \centering
  \begin{minipage}[t]{0.49\textwidth}
    \centering
    \includegraphics[width=\linewidth]{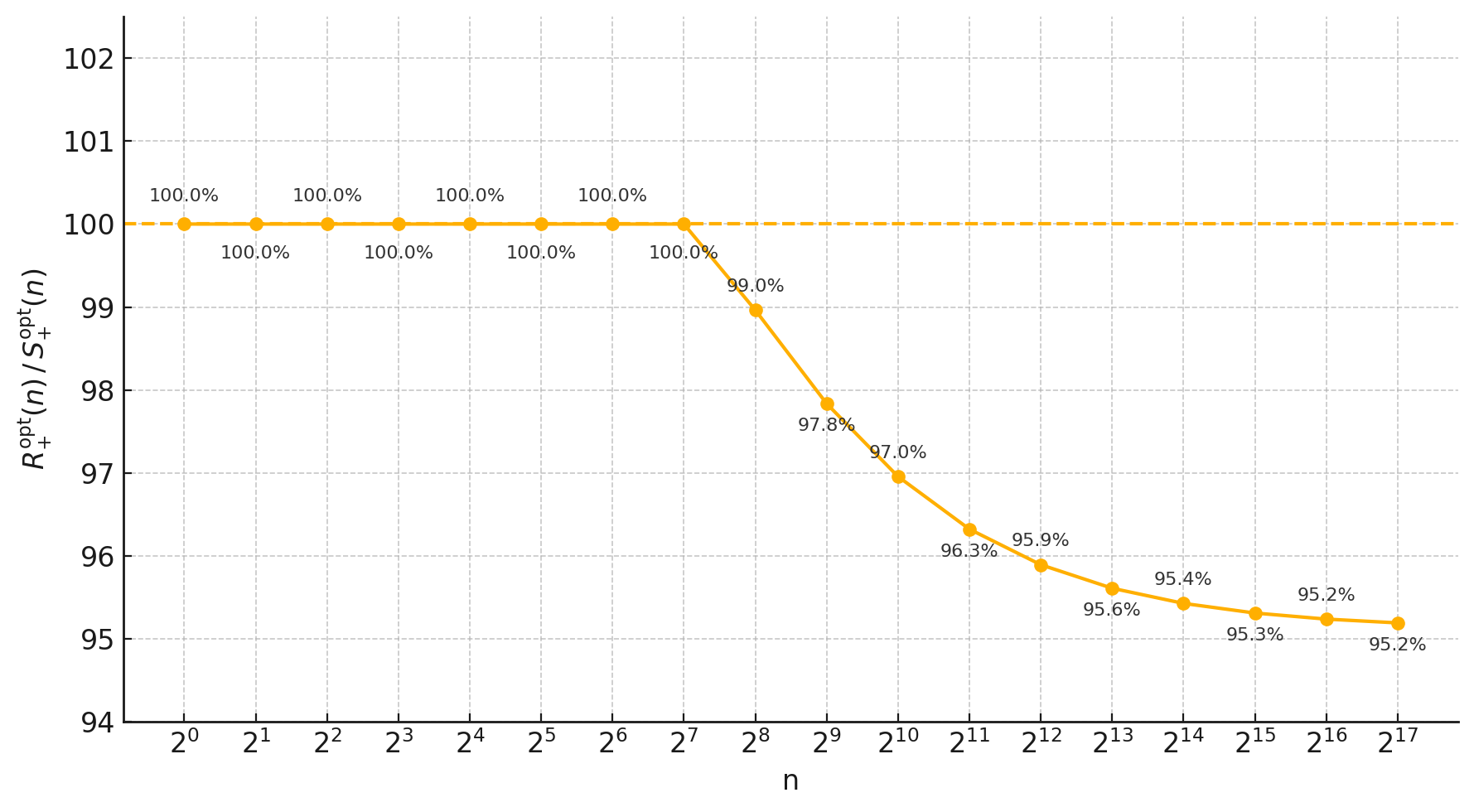}
    Ratio of $R^{opt}_{+}(n)$ to $S^{opt}_{+}(n)$.
  \end{minipage}
  \hfill
  \begin{minipage}[t]{0.49\textwidth}
    \centering
    \includegraphics[width=\linewidth]{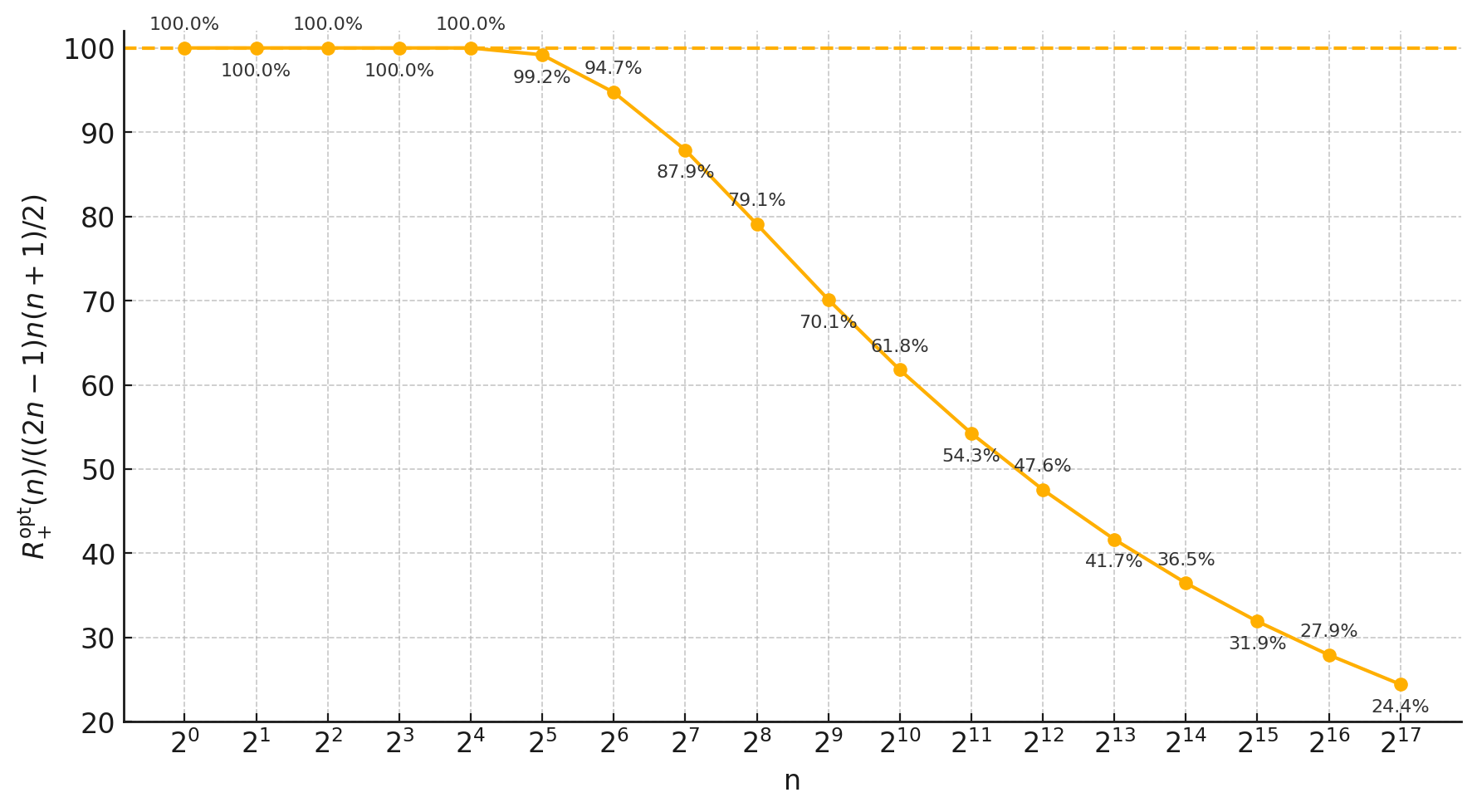}
    Ratio of $R^{opt}_{+}(n)$ to $(2n-1)n(n+1)/2$.
  \end{minipage}
  \caption{Comparison of algorithms with optimal cutoffs i.e. for small enough matrices in recursion switch to the algorithm with least operations. RXTX outperforms naive algorithm for $n \geq 32$ and SotA for $n \geq 256$.}
  \label{fig:operations_opt}
\end{figure}

\newlength{\firstcol}               % width of the first column
\setlength{\firstcol}{46mm}        % adjust if you need a wider/narrower gap
\newcommand{\pair}[2]{\makebox[\firstcol][l]{$#1$} $#2$}

% stage 1
\begin{algorithm}[H]
\caption{First stage of optimized addition scheme. The number of additions is reduced from $77$ to $53$.}\label{alg:stage1}
\begin{algorithmic}[1]
\State \textbf{Input:}  $X_1, X_2, ..., X_{16}$

\State \textbf{Output:} Left elements $L_1, ..., L_{26}$ and right elements $R_1,..., R_{26}$ of multiplications $m_1, ... m_{26}$

\State $y_1 \gets X_{13} - X_{14}$
\State $y_2 \gets X_{12} - X_{10}$
\State $w_1  \gets X_2 + X_4 - X_8$
\State $w_2  \gets X_1 - X_5 - X_6$
\State $w_3  \gets X_6 + X_7$
\State $w_4  \gets X_{14} + X_{15}$
\State $w_5  \gets y_2 + X_{16}$
\State $w_6  \gets X_{10} + X_{11}$
\State $w_7  \gets X_9 + y_1$
\State $w_8  \gets X_9 - X_8$
\State $w_9  \gets X_7 - X_{11}$
\State $w_{10} \gets X_6 - X_7$
\State $w_{11} \gets X_2 - X_3$
\State \pair{L_{1} \gets -w_1 + X_3}{R_{1}\gets X_8 + X_{11}}
\State \pair{L_{2} \gets w_2 + X_7}{R_{2}\gets X_{15} + X_5}
\State \pair{L_{3} \gets -X_2+X_{12}}{R_{3}\gets w_5}
\State \pair{L_{4} \gets X_9-X_6}{R_{4}\gets w_7}
\State \pair{L_{5} \gets X_2+X_{11}}{R_{5}\gets X_{15}-w_3}
\State \pair{L_{6} \gets X_6+X_{11}}{R_{6}\gets w_3-X_{11}}
\State \pair{L_{7} \gets X_{11}}{R_{7}\gets w_3}
\State \pair{L_{8} \gets X_2}{R_{8}\gets w_3-w_4+w_5}
\State \pair{L_{9} \gets X_6}{R_{9}\gets w_7-w_6+w_3}
\State \pair{L_{10} \gets w_1-X_3+X_7+X_{11}}{R_{10}\gets X_{11}}
\State \pair{L_{11} \gets X_5+w_{10}}{R_{11}\gets X_5}
\State \pair{L_{12} \gets w_{11}+X_4}{R_{12}\gets X_8}
\State \pair{L_{13} \gets -w_2+X_3-w_9}{R_{13}\gets X_{15}}
\State \pair{L_{14} \gets -w_2}{R_{14}\gets w_7+w_4}
\State \pair{L_{15} \gets w_1}{R_{15}\gets w_6+w_5}
\State \pair{L_{16} \gets X_1-X_8}{R_{16}\gets X_9-X_{16}}
\State \pair{L_{17} \gets X_{12}}{R_{17}\gets -y_2}
\State \pair{L_{18} \gets X_9}{R_{18}\gets y_1}
\State \pair{L_{19} \gets -w_{11}}{R_{19}\gets -X_{15}+X_7+X_8}
\State \pair{L_{20} \gets X_5+w_8}{R_{20}\gets X_9}
\State \pair{L_{21} \gets X_8}{R_{21}\gets X_{12}+w_8}
\State \pair{L_{22} \gets -w_{10}}{R_{22}\gets X_5+w_9}
\State \pair{L_{23} \gets X_1}{R_{23}\gets X_{13}-X_5+X_{16}}
\State \pair{L_{24} \gets -X_1+X_4+X_{12}}{R_{24}\gets X_{16}}
\State \pair{L_{25} \gets X_9+X_2+X_{10}}{R_{25}\gets X_{14}}
\State \pair{L_{26} \gets X_6+X_{10}+X_{12}}{R_{26}\gets X_{10}}
\end{algorithmic}
\end{algorithm}

% stage 2
\begin{algorithm}[H]
\caption{Second stage of optimized addition scheme. The number of additions is reduced from $62$ to $47$.}\label{alg:stage2}
\begin{algorithmic}[1]
\State \textbf{Input:} $m_1, m_2, ..., m_{26}$ and $s_1, ... s_8$.
\State \textbf{Output:} Entries $C_{ij}$ using 47 additions.
% ------------------ auxiliary sums (z_i) ----------------------------------
\State $z_1 \gets m_7  \,-\, m_{11} \,-\, m_{12}$
\State $z_2 \gets m_1  \,+\, m_{12} \,+\, m_{21}$
\State $z_3 \gets m_3  \,+\, m_{17} \,-\, m_{24}$
\State $z_4 \gets m_2  \,+\, m_{11} \,+\, m_{23}$
\State $z_5 \gets m_5  \,+\, m_7   \,+\, m_8$
\State $z_6 \gets m_4  \,-\, m_{18} \,-\, m_{20}$
\State $z_7 \gets m_6  \,-\, m_7   \,-\, m_9$
\State $z_8 \gets m_{17} \,+\, m_{18}$
  % ------------------ target coefficients ----------------------------------
\State $C_{11} \gets s_1 + s_2 + s_3 + s_4$
\State $C_{12} \gets m_2 \,-\, m_5 \,-\, z_1 \,+\, m_{13} \,+\, m_{19}$
\State $C_{13} \gets z_2 \,+\, z_3 \,+\, m_{15} \,+\, m_{16}$
\State $C_{14} \gets z_4 \,-\, z_3 \,-\, z_5 \,+\, m_{13}$
\State $C_{22} \gets m_1 \,+\, m_6 \,-\, z_1 \,+\, m_{10} \,+\, m_{22}$
\State $C_{23} \gets z_2 \,-\, z_6 \,+\, z_7 \,+\, m_{10}$
\State $C_{24} \gets z_4 \,+\, z_6 \,+\, m_{14} \,+\, m_{16}$
\State $C_{33} \gets m_4 \,-\, z_7 \,-\, z_8 \,+\, m_{26}$
\State $C_{34} \gets m_3 \,+\, z_5 \,+\, z_8 \,+\, m_{25}$
\State $C_{44} \gets s_5 + s_6 + s_7 + s_8$
\end{algorithmic}
\end{algorithm}

\subsection{Runtime of RXTX}
\begin{figure}[H]
    \centering
    \includegraphics[width=0.8\linewidth]{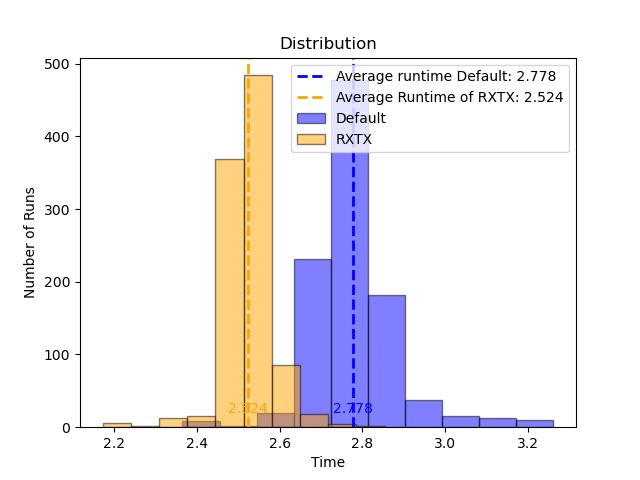}
    \caption{The average runtime for RXTX is $2.524$s, which is 9\% faster than average runtime of specific BLAS routine $2.778$s. RXTX was faster in $99\%$ of the runs.}
    \label{fig:rxtx_runtime}
\end{figure}

We verify that RXTX gives speed-up in practice for large sizes of matrix $X$. Figure \ref{fig:rxtx_runtime} shows histogram of runtimes in the following setup:
\begin{itemize}
    \item $6144 \times 6144$ dense matrix is sampled $1000$ times with random normal $\mathcal{N}(0, 1)$ entries. Here $6144 =3\cdot 2^{12}$. 
    \item RXTX is implemented as depth-1 recursion i.e. we directly use BLAS routines to compute $26$ general matrix multiplications and $8$ symmetric products of matrices of size $1536\times 1536$.
    \item Default is direct call of BLAS-3 routine \citep{blas} for $XX^{t}$.
    \item single thread CPU 10th Gen Intel Core i7-10510U Processor, 1.8 GHz 4 cores.
\end{itemize}
We didn't perform search for the smallest matrix size where RXTX outperforms other methods since runtime is highly sensitive to hardware, computation graph organization, and memory management. Figure \ref{fig:operations_opt} suggests that RXTX can be faster than recursive Strassen for $n \geq 256$.

\section{Discovery Methodology}
\subsection{Description of RL-guided Large Neighborhood Search}
\label{methodology}
In this section we briefly present our methodology. Full methodology with other discovered accelerations will be described in \citep{rybin2025}. We combine RL-guided Large Neighborhood Search \citep{wu2021, addanki2020} with a two–level MILP pipeline: 
\begin{enumerate}
    \item The RL agent proposes a (potentially redundant) set of rank-1 bilinear products;
    \item MILP-A exhaustively enumerates tens of thousands of linear relations between these candidate rank-1 bilinear products and target expressions;
    \item MILP-B then selects the smallest subset of products whose induced relations cover every target expression of $XX^{t}$.
\end{enumerate} 
The loop iterates under a Large Neighborhood Search regime. One way to view this pipeline is a simplification of AlphaTensor RL approach \citep{fawzi2022}: instead of sampling tensors from $\mathbb{R}^{n^2} \otimes \mathbb{R}^{n^2} \otimes \mathbb{R}^{n^2}$, we sample candidate tensors from $\mathbb{R}^{n^2} \otimes \mathbb{R}^{n^2}$ and let the MILP solver find optimal linear combinations of sampled candidates. This reduces the action space by a factor $\times 1000000$.

\subsection{Example: matrix times transpose algorithm search for 2-by-2 matrix}
Consider the example for $2\times 2$ matrix $X$. We want to perform the computation of $XX^{t}$:
$$\begin{pmatrix}
    x_1 & x_2 \\
    x_3 & x_4
\end{pmatrix} \cdot \begin{pmatrix}
    x_1 & x_3 \\
    x_2 & x_4
\end{pmatrix} = \begin{pmatrix}
    x_{1}^{2} + x_{2}^{2} & x_{1}x_{3} + x_{2}x_{4} \\
    x_{1}x_{3} + x_{2}x_{4} & x_{3}^{2} + x_{4}^{2}
\end{pmatrix}
$$
We identify $3$ target expressions
$$T = \{x_{1}^{2} + x_{2}^{2}, x_{3}^{2} + x_{4}^{2}, x_{1}x_{3} + x_{2}x_{4}\}.$$
We randomly sample thousands of products $p_1, ..., p_m$, each one given by
$$\left(\sum_{i=1}^{4}\alpha_ix_i \right)\cdot \left(\sum_{j=1}^{4}\beta_jx_j\right)$$
with $\alpha_i, \beta_j \in \{-1, 0, +1\}$ chosen by RL policy $\pi_{\theta}$. MILP-A enumerates ways to write target expressions from $T$ as linear combinations of sampled products $\sum \gamma_ip_i$. MILP-B selects minimal number of sampled products such that every target expression can be obtained as their linear combination. Key observation is that MILP-A and MILP-B are rapidly solvable with solvers like Gurobi \citep{gurobi}.

\section{Conclusion}

In this work, we presented RXTX, a new algorithm for computing $XX^{t}$ for $X\in \mathbb{R}^{n\times m}$.  RXTX uses 5\% fewer number of  multiplications and 5\% fewer total number of operations (additions and multiplications) than SotA algorithms. Importantly, the speedup not only holds asymptotically for large matrices with $n \rightarrow \infty$, but also for small matrices.

\section*{Acknowledgements}
The work of Z.-Q. Luo was supported by the Guangdong Major
Project of Basic and Applied Basic Research (No.2023B0303000001), the Guangdong Provincial Key Laboratory of Big Data Computing, and the National Key Research and Development Project under grant 2022YFA1003900.

\bibliographystyle{abbrvnat}
\bibliography{reference.bib}

\end{document}